\documentclass[journal]{IEEEtran}
\usepackage{amsmath,amsfonts}
\usepackage{algorithmic}
\usepackage{algorithm}
\usepackage{array}
\usepackage[caption=false,font=normalsize,labelfont=sf,textfont=sf]{subfig}
\usepackage{textcomp}
\usepackage{stfloats}
\usepackage{url}
\usepackage{verbatim}
\usepackage{graphicx}
\usepackage{cite}
\usepackage{tabularx,booktabs,multirow,multicol,rotating,bbding}
\usepackage{pifont}
\usepackage[super]{nth}

\begin{document}

\title{Improving Acoustic Scene Classification with City Features}

\author{Yiqiang Cai, Yizhou Tan,~\IEEEmembership{Student Member,~IEEE}, Shengchen Li,~\IEEEmembership{Senior Member,~IEEE}, \\Xi Shao,~\IEEEmembership{Senior Member,~IEEE}, and Mark D. Plumbley,~\IEEEmembership{Fellow,~IEEE}
\thanks{This paper was produced by the IEEE Publication Technology Group. They are in Piscataway, NJ.}
\thanks{Manuscript received April 19, 2021; revised August 16, 2021.}}

\markboth{Journal of \LaTeX\ Class Files,~Vol.~14, No.~8, August~2021}%
{Shell \MakeLowercase{\textit{et al.}}: A Sample Article Using IEEEtran.cls for IEEE Journals}


\maketitle

\begin{abstract}
Acoustic scene recordings are often collected from a diverse range of cities. Most existing acoustic scene classification (ASC) approaches focus on identifying common acoustic scene patterns across cities to enhance generalization. However, the potential acoustic differences introduced by city-specific environmental and cultural factors are overlooked. In this paper, we hypothesize that the city-specific acoustic features are beneficial for the ASC task rather than being treated as noise or bias. To this end, we propose City2Scene, a novel framework that leverages city features to improve ASC. Unlike conventional approaches that may discard or suppress city information, City2Scene transfers the city-specific knowledge from pre-trained city classification models to scene classification model using knowledge distillation. We evaluate City2Scene on three datasets of DCASE Challenge Task 1, which include both scene and city labels. Experimental results demonstrate that city features provide valuable information for classifying scenes. By distilling city-specific knowledge, City2Scene effectively improves accuracy across a variety of lightweight CNN backbones, achieving competitive performance to the top-ranked solutions of DCASE Challenge in recent years.
\end{abstract}


\section{Introduction}
\label{sec:intro}
\IEEEPARstart{A}{coustic} scene classification (ASC) \cite{barchiesi2015acoustic} is a fundamental task in the field of machine listening, with applications ranging from environmental monitoring to personalized user experiences in smart devices. The objective of ASC is to categorize short audio clips into pre-defined scene classes such as park, airport, or shopping mall. To enhance acoustic diversity, audio for each scene class is typically recorded across various geographical regions during the data collection phrase. Consequently, these audio samples are annotated with both scene and region labels. In the majority of ASC datasets \cite{Mesaros2018, Heittola2020, bai2024description}, the region labels correspond to cities, rather than smaller units like neighborhoods or broader regions like countries. Cities, as relatively independent geographical and administrative entities, provide a wide range of acoustic scenes. Moreover, cities often possess unique cultural and environmental characteristics that are reflected in their acoustic scenes, offering data variability for the development of robust ASC systems.

However, acoustic variations between cities are mainly treated as noise or bias in ASC datasets. The Detection and Classification of Acoustic Scenes and Events (DCASE) challenge pioneered the introduction of a multi-city dataset for ASC, namely TUT Urban Acoustic Scenes 2018 dataset \cite{Mesaros2018}, which comprises 24 hours of high-fidelity audio captured across six European cities. Subsequent versions of the DCASE datasets include recordings from additional cities, with some cities exclusively appear in the evaluation set, potentially leading to domain shift issues \cite{Heittola2020}. The problem of domain shift across cities was further emphasized in the ICME 2024 Grand Challenge \cite{bai2024description} by including more unseen cities in the evaluation set, increasing the difficulty of generalization. These data challenges have inspired researchers to focus on identifying common acoustic scene patterns across different cities. The subsequent solutions include data augmentation \cite{hu2021two}, domain adaptation \cite{singh2021prototypical}, transfer learning \cite{liu2021cross} and loss configuration \cite{yan2024semi}.

\begin{figure}
\centering
\includegraphics[width=\linewidth]{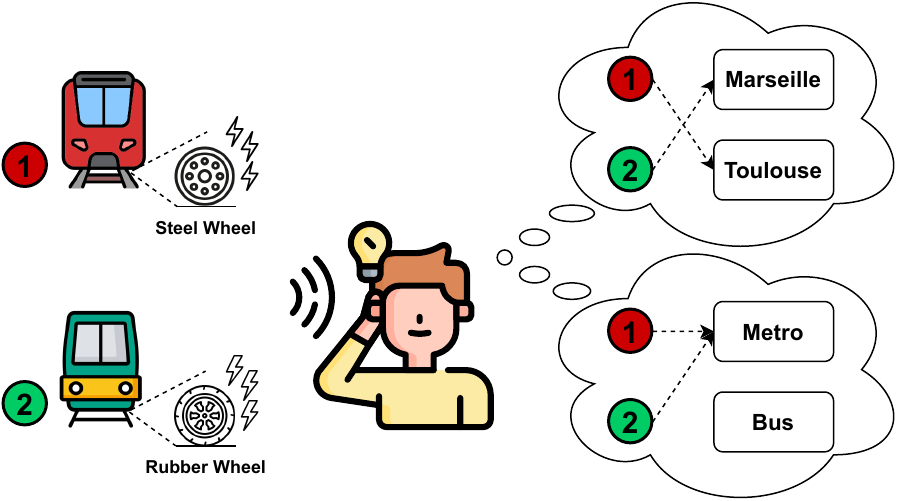} 
\caption{The sound of metro wheels can tell us apart cities (\textit{Toulouse} and \textit{Marseille}), and also scenes (\textit{metro} and \textit{bus}).}
\label{fig:metro-example}
\end{figure}

Despite significant improvements in generalization, the role of city-specific features in ASC remains largely underexplored. In real world, acoustic features of environmental and cultural differences unique to cities could offer valuable clues for distinguishing scenes. For example, as illustrated in Figure \ref{fig:metro-example}, the sound of metro wheels could serve as an indicator to differentiate not only \textit{Marseille} from \textit{Toulouse} but also \textit{metro} from \textit{bus}.  Several previous works have explored the introduction of city information to ASC. Bear et al. \cite{bear2019city} proposed a multi-task learning (MTL) framework which shares feature extractor for scene and city classification. However, MTL may create conflicting gradients between the main task and the auxiliary task, leading the model to suppress scene features. In contrast, Tan et al. \cite{tan2024acoustic} aimed to disentangle city features from acoustic scenes to address the domain shift challenge, while the disentanglement ignored the potential connection between cities and scenes. To explore the benefits of city features in ASC, we conduct a preliminary experiment. Specifically, the features extracted from pre-trained city classification models are used to classify acoustic scenes. Experimental results show that the performance of city features in scene classification tasks is highly comparable to that of scene features, demonstrating city features contain valuable information to differentiate scenes.

Motivated by the findings, we introduce City2Scene, a novel framework that leverages audio features with city-related information to improve the classification accuracy of ASC models. Initially, city labels are used to train a city classification model. The frozen city features are then transformed into city-to-scene soft targets, reflecting city characteristics in soft scene labels. By using a general knowledge distillation framework \cite{Schmid2023workshop}, the city-specific knowledge is integrated to the training process of ASC model, transferring the unique acoustic characteristics of various cities. To validate our approach, we perform extensive experiments using several state-of-the-art CNN backbones on three ASC datasets that include recordings from various cities. Comparing to existing methods, experimental results show that City2Scene consistently improves the performance of ASC models on three datasets, and the use of knowledge distillation avoids introducing additional computational overhead in the inference stage. Furthermore, the combination of City2Scene and TF-SepNet achieves the best accuracy of 63.6\% on the TAU22 dataset \cite{Heittola2020}, surpassing the top-ranked systems in the DCASE challenges of 2022 and 2023. This study highlights the importance of city-specific information in ASC and paves the way for future research in location-assisted ASC.

\begin{figure*}[t]
\centering
\includegraphics[width=1.0\linewidth]{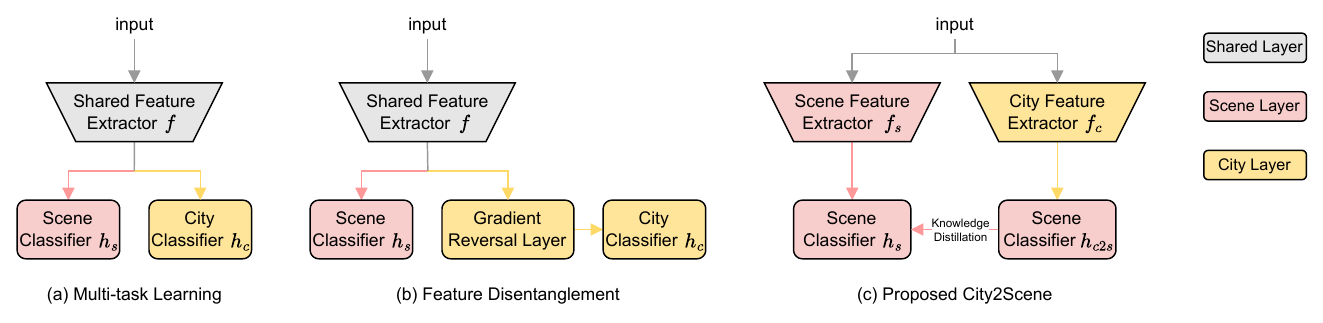} 
\caption{Comparison of frameworks for incorporating city information into acoustic scene classification. (a) \textbf{Multi-task learning} simultaneously optimizes scene and city classification objectives with shared representations. (b) \textbf{Feature disentanglement} introduces a domain-adversarial loss via a gradient reversal layer (GRL) to separate city-related and scene-related features. (c) The \textbf{proposed City2Scene} framework distills city-specific knowledge from the city classification model to the scene classification model.}
\label{fig:methods}
\end{figure*}

To summarize the main contributions of this work, we:

\begin{itemize}
    \item demonstrate that city classification models capture rich acoustic features that are beneficial for ASC.
    \item propose City2Scene, a novel framework that improves ASC by incorporating city features through knowledge distillation.
    \item validate City2Scene across various state-of-the-art backbones on three representative ASC datasets, showing consistent improvements over existing methods.
\end{itemize}

The remainder of this paper is organized as follows. Section \ref{sec:related_work} reviews related work on ASC with city information and knowledge distillation. Section \ref{sec:proposed_method} introduces the proposed City2Scene framework. Section \ref{sec:experiment} details the experimental setup, including datasets, model backbones, and implementation details. Section \ref{sec:results} presents the results and analysis. Section \ref{sec:discussion} discusses the benefits and limitations of city features for ASC. Finally, Section \ref{sec:conclusion} concludes the paper and outlines potential future directions.

\section{Related Work}
\label{sec:related_work}
In this section, we review two major research directions that are closely related to our proposed method: the use of city information to enhance acoustic scene classification (ASC), and the application of knowledge distillation to improve ASC models. We discuss the progress and limitations of these approaches in the following subsections.

\subsection{ASC with City Information}
\label{ssec:related_methods}
ASC has seen substantial progress over the past decade, particularly through the annual DCASE Challenge\footnote{\url{https://dcase.community/}}, which has continually expanded datasets in terms of size, diversity, and recording conditions. However, it remains challenging to extract discriminative features due to the short duration (often 1-10 seconds) of audio clips \cite{Martin-Morato2022}, the limited quantity of labeled training data \cite{Schmid2024a}, and domain shift issues \cite{Heittola2020}.

To overcome these challenges, prior research has explored the use of auxiliary domain-specific information, such as metadata about the recording city, to enhance ASC performance. One effective approach is multi-task learning (MTL) \cite{Caruana1998}, where a shared feature extractor is trained by simultaneously classifying both scenes and auxiliary task classes. For instance, Bear et al. \cite{bear2019city} train a shared feature extractor with two classifiers to classify city labels alongside scene labels as illustrated in Figure \ref{fig:methods}(a). Let $f(\cdot)$ be a shared feature extractor applied to the input acoustic feature $x$, and let $h_s(\cdot)$ and $h_c(\cdot)$ denote the scene and city classifiers, respectively. Formally, the total loss for MTL can be defined as:
\begin{equation}
    \mathcal{L}_{\text{MTL}} = \mathcal{L}_{\text{scene}}(h_s(f(x)), y_s)+\alpha\cdot \mathcal{L}_{\text{city}}(h_c(f(x)), y_c),
\end{equation}
where $\mathcal{L}_{\text{scene}}$ is the cross-entropy loss for scene classification, $\mathcal{L}_{\text{city}}$ is the auxiliary task loss (i.e., city classification), and $\alpha$ is a weight controlling the influence of the auxiliary task. MTL encourages the model to extract shared and complementary features between tasks. However, MTL relies on careful task balancing, and the auxiliary task may dominate learning if not appropriately weighted.

Another approach to leveraging city information is feature disentanglement (FD), which aims to extract domain-invariant features by discouraging the model from encoding domain-specific information in the shared representation. Tan et al. \cite{tan2024acoustic} proposes to remove city biases from learned features through adversarial training. As show in Figure \ref{fig:methods}(b), a gradient reversal layer (GRL) \cite{ganin2016domain} is inserted before the city classifier. The model is trained to minimize the loss of scene classification while simultaneously making the city classifier unable to predict the cities from the same features. The GRL reverses the gradient by multiplying $-\beta$ during backpropagation, encouraging the feature extractor to be domain-invariant. The total objective function for FD is:
\begin{equation}
    \mathcal{L}_{\text{FD}} = \mathcal{L}_{\text{scene}}(h_s(f(x)), y_s) - \beta \cdot \mathcal{L}_{\text{city}}(h_c(f(x)), y_c).
\end{equation}
While FD leads to domain-invariant representations, it may also discard useful information if domain feature overlaps with scene characteristics.

Despite the limitations, both approaches highlight the potential of leveraging city information to improve ASC. Rather than discarding or suppressing city features, our work proposes to transfer the city features to scene classifiers through knowledge distillation as shown in Figure \ref{fig:methods}(c).

\subsection{Knowledge Distillation}
In practical applications, ASC systems often face strict constraints on model complexity due to limited computing resources on edge devices such as mobile phones, embedded boards, or IoT sensors \cite{Heittola2020}. To meet the constraints while maintaining performance, researchers have developed a variety of techniques, including quantization \cite{Martin2021}, pruning \cite{singh22_interspeech}, and the design of lightweight neural network architectures \cite{koutini2019receptive, kim21l_interspeech, cai2024tf}. However, these methods typically involve trade-offs between complexity and accuracy.

Knowledge distillation (KD) \cite{hinton2015distilling} has emerged as a powerful technique to reduce model complexity while preserving accuracy. In the commonly used KD framework, a large, accurate teacher model transfers its knowledge to a smaller, efficient student model using soft labels. Let $\mathbf{z}^{(t)}$ and $\mathbf{z}^{(s)}$ be the logits from the teacher and student models, respectively. The soft labels are obtained by applying a temperature-scaled softmax to the logits:
\begin{equation}
    \delta(z_i, \tau) = \frac{\exp(z_i / \tau)}{\sum_j \exp(z_j / \tau)}.
\end{equation}
Here, $\tau > 1$ is the temperature parameter that smooths the output distribution. Formally, the distillation loss is defined as the Kullback-Leibler divergence (KLD) between the softmax outputs of teacher model and student model:

\begin{equation}
    \mathcal{L}_{\text{KD}} = \tau^2 \cdot \mathcal{L}_{\text{KLD}}(\delta(\mathbf{z}^{(t)}, \tau), \delta(\mathbf{z}^{(s)}, \tau)).
\end{equation}

The total loss typically combines the distillation loss with the standard cross-entropy loss $\mathcal{L}_{\text{CE}}$ with true labels:

\begin{equation}
    \mathcal{L} = \lambda \cdot \mathcal{L}_{\text{CE}} + (1 - \lambda) \cdot \mathcal{L}_{\text{KD}},
\end{equation}

\noindent
where $\lambda \in [0, 1]$ balances the contribution of each loss component.

In the context of ASC, KD has shown impressive results in recent DCASE Challenges \cite{Schmid2023workshop, schmid2022knowledge, han2024data}. Schmid et al. \cite{schmid2022knowledge} demonstrates that Transformer-based models, although typically too large for deployment, can be used as effective teachers for CNN-based students, improving accuracy while maintaining efficiency. Similarly,  Schmid et al. \cite{Schmid2023workshop} explores ensemble knowledge distillation, where logits of both Transformer and CNN teachers are averaged as the teacher knowledge. Our work builds upon the successes of KD for ASC by further exploring task-specific distillation, where knowledge from the city classification teacher is transferred to enhance scene classification.

\begin{figure}
\centering
\includegraphics[width=\linewidth]{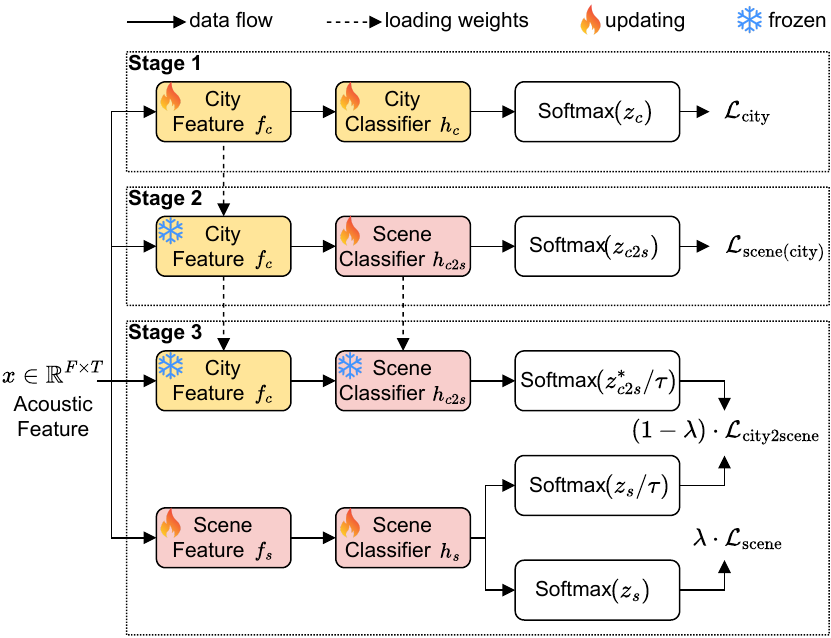} 
\caption{Illustration of the proposed City2Scene framework. The input feature $x$ is in $\mathbb{R}^{F \times T}$, where $F, T$ respectively denotes frequency and time dimensions. \textbf{Stage 1:} Classifying cities. \textbf{Stage 2:} Classifying scenes using frozen city features. \textbf{Stage 3:} Distilling city knowledge to ASC model.}
\label{fig:city2scene}
\end{figure}

\section{Proposed Method}
\label{sec:proposed_method}
In this section, we present the proposed City2Scene framework in detail. As shown in Figure \ref{fig:city2scene}, City2Scene consists of three training stages. Section \ref{ssec:classifying} focuses on the first two stages: training a city classification model and then using the frozen city features to classify acoustic scenes. This section also serves as a preliminary experiment to explore the effectiveness of city features for the ASC task. Section \ref{ssec:ditilling} describes the third stage that distills city-specific knowledge into the ASC model. Finally, Section \ref{ssec:fusing} introduces the fusion of knowledge from the city teacher and the scene teacher during knowledge distillation.

\subsection{Classifying Acoustic Scenes using City Features}
\label{ssec:classifying}
The first stage in the City2Scene framework is to train a city classification model, which consists of a city feature extractor and a city classifier. The city feature extractor $f_c(\cdot)$ is used to process the input acoustic feature $x \in\mathbb{R}^{F\times T}$, outputting a feature embedding that encapsulates city-specific information. The feature is then passed through a city classifier $h_c(\cdot)$, usually a fully connected (FC) layer, followed by a softmax function $\delta(\cdot)$, to predict the city class probabilities. The training process minimizes the city classification loss $\mathcal{L}_{\mathrm{city}}$, defined as:

\begin{equation}
    z_c = h_c(f_c(x)),
\end{equation}
\begin{equation}
    \mathcal{L}_{\mathrm{city}} = \mathcal{L}_{\mathrm{CE}}(\delta(z_c), y_c),
\end{equation}

\noindent
where $\mathcal{L}_{\mathrm{CE}}$ is the cross-entropy loss, $z_c$ represents the logits of city classes, and $y_c$ is the ground truth city label.

In the second stage, as illustrated in Figure \ref{fig:city2scene}, the parameters of city feature extractor will be frozen once the city classification model is trained. The intermediate features produced by the frozen city feature extractor $f^{*}_c(\cdot)$ represent high-level embeddings of the acoustic characteristics associated with different cities. These frozen city features are passed to a scene classifier $h_{c2s}$. The loss function for classifying acoustic scenes using city features $\mathcal{L}_{\mathrm{scene(city)}}$ is defined as:

\begin{equation}
    z_{c2s} = h_{c2s}(f_c^*(x)),
\end{equation}
\begin{equation}
    \mathcal{L}_{\mathrm{scene(city)}} = \mathcal{L}_{\mathrm{CE}}(\delta(z_{c2s}), y_s),
\end{equation}

\noindent
where $z_{c2s}$ is the city-to-scene logits produced by the scene classifier, and $y_s$ is the ground truth scene label. Furthermore, after the first two training stages, we can also evaluate the effectiveness of city features for the ASC task by examining the performances of city classifier and scene classifier.

\subsection{Distilling City Knowledge to ASC Model}
\label{ssec:ditilling}
The third stage of City2Scene focuses on transferring the city-specific knowledge extracted from the pre-trained combination of city feature extractor and scene classifier to the ASC model through knowledge distillation. The frozen city feature extractor $f^{*}_c(\cdot)$ and scene classifier $h^{*}_{c2s}(\cdot)$ provide city-to-scene logits $z_{c2s}^*$, which capture information about how city characteristics contribute to scene classification. These logits are used as the ``teacher" to guide the training of the ASC model. The ASC model serves as the ``student" in the knowledge distillation process, which receives acoustic inputs and learns to replicate the predictions of the teacher model while simultaneously optimizing for scene classification accuracy. Finally, the ASC student model is trained with a combination of two loss functions, distillation loss $\mathcal{L}_{\mathrm{city2scene}}$ and scene classification loss $\mathcal{L}_{\mathrm{scene}}$.

Scene classification loss $\mathcal{L}_{\mathrm{scene}}$ ensures that the student model predicts the correct scene labels. It is defined as:

\begin{equation}
    \mathbf{z}^{(s)} = z_s = h_s(f_s(x)),
\end{equation}
\begin{equation}
\label{eq:scene}
    \mathcal{L}_{\mathrm{scene}} = \mathcal{L}_{\mathrm{CE}}(\delta(z_s), y_s),
\end{equation}

\noindent
where $f_s$ and $h_s$ are respectively scene feature extractor and scene classifier, $z_s$ represents the logits for scene classification, $\mathbf{z}^{(s)}$ denotes the logits of the student model, $\delta(z_s)$ is the softmax output, and $y_s$ is the ground truth scene label.

Distillation loss $\mathcal{L}_{\mathrm{city2scene}}$ enables the ASC model to mimic the softened predictions of the teacher model. It is defined as: 

\begin{equation}
    \mathbf{z}^{(t)} = z^{*}_{c2s} = h^{*}_{c2s}(f^{*}_c(x)),
\end{equation}
\begin{equation}
\label{eq:city2scene}
    \mathcal{L}_{\mathrm{city2scene}} = \tau^2\cdot\mathcal{L}_{\mathrm{KLD}}(\delta(\mathbf{z}^{(s)}, \tau),  \delta(\mathbf{z}^{(t)}, \tau)),
\end{equation}

\noindent
where $\mathcal{L}_{\mathrm{KLD}}$ represents the Kullback-Leibler divergence, $\tau$ is the temperature scaling factor to soften the probability distributions, and $\mathbf{z}^{(t)}$ is the logits of the teacher model. Following the success of ensembling teachers in \cite{Schmid2023workshop}, we additionally create teacher model ensembles to distill diverse knowledge from different architectures, where $\mathbf{z}^{(t)} = \frac{1}{N} \sum_{i=1}^N z^{*}_{c2s_{(i)}}$ and $N$ represents the number of teacher architectures.

The overall training objective for the ASC model is:

\begin{equation}
\label{eq:loss}
    \mathcal{L} = \lambda\cdot\mathcal{L}_{\mathrm{scene}} + (1-\lambda)\cdot\mathcal{L}_{\mathrm{city2scene}},
\end{equation}

\noindent
where $\lambda$ controls the trade-off between label and distillation loss. By combining the losses, the ASC model learns not only to classify scenes but also to integrate the city-specific knowledge encoded by the teacher model.

\begin{figure}
\centering
\includegraphics[width=\linewidth]{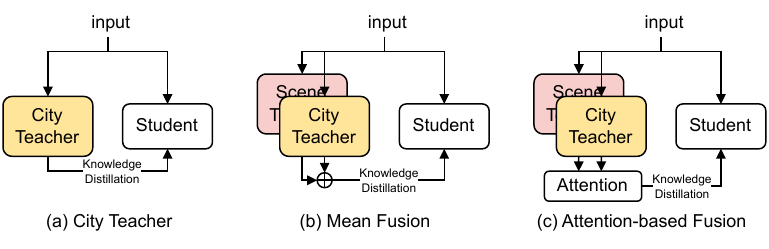} 
\caption{Knowledge fusion from city and scene teachers. \textbf{(a)} Distilling city knowledge to ASC model. \textbf{(b)} Mean fusion of city and scene knowledge. \textbf{(c)} Weighted knowledge fusion with attention layer.}
\label{fig:teachers}
\end{figure}

\subsection{Fusing City and Scene Knowledge}
\label{ssec:fusing}
While prior works \cite{Schmid2023workshop,han2024data} focused on distilling knowledge from multiple scene teachers with diverse architectures or feature extraction strategies, we extend this direction by exploring the fusion of knowledge from different task teachers, i.e. city and scene teachers, as illustrated in Figure \ref{fig:teachers}. The intuition is that city models capture environmental and cultural acoustic patterns, while scene models focus on task-specific patterns. In addition to simply distilling city-specific knowledge in Figure \ref{fig:teachers}(a), we implement two strategies for teacher fusion. The first is a simple mean fusion of logits from city and scene teachers, as in Figure \ref{fig:teachers}(b):

\begin{equation}
    \mathbf{z}^{(t)} = z_{\mathrm{mean}} = \frac{1}{N} \sum_{i=1}^N \frac{z_{c2s_{(i)}}^* + z_{s_{(i)}}}{2},
\end{equation}

\noindent
where $N$ denotes the number of teacher architectures, $z_{c2s_{(i)}}$ and $z_{s_{(i)}}$ are respectively logits generated by the $i$-th city teacher and scene teacher.

The second strategy introduces a trainable attention weight $\gamma$ to balance contributions from city and scene teachers, as in Figure \ref{fig:teachers}(c):

\begin{equation}
\label{eq:attn}
    \gamma_{(i)} = \delta(W[z_{c2s_{(i)}}^* \,\|\, z_{s_{(i)}}]),
\end{equation}

\begin{equation}
\label{eq:attn_teacher}
    \mathbf{z}^{(t)} = z_{\mathrm{attn}} = \frac{1}{N} \sum_{i=1}^N (\gamma_{(i)} \cdot z_{c2s_{(i)}} ^*+ (1-\gamma_{(i)}) \cdot z_{s_{(i)}}),
\end{equation}

\noindent
where $\,\|\,$ denotes concatenation, $W$ is a learnable linear layer that transforms the concatenated logits into attention scores.

\section{Experiment}
\label{sec:experiment}
This section presents the experimental validation of the proposed City2Scene framework for ASC. Details of the datasets, model backbones, and training setup are described in the following subsections.

\subsection{Datasets}
The experiments are conducted on the DCASE Challenge Task 1 datasets, namely the TAU Urban Acoustic Scene 2020 / 2022 / 2024 Mobile development datasets (TAU20 / TAU22 / TAU24) \cite{Heittola2020}, where each audio sample is annotated with both a city label and a scene label. All three datasets have ten city classes and ten scene classes, with audio samples approximately evenly distributed across cities and scenes. Both classes of city and scene remain consistent between training set and test set. Furthermore, audio in all datasets were recorded using various devices, some exclusive to the test set, resulting in difficulty for generalization. The primary difference between TAU20 and TAU22 lies in the duration of the audio clips: TAU20 contains 10-second clips, whereas TAU22 comprises 1-second clips. The shorter clip length introduces more challenges for ASC models to capture acoustic patterns. TAU24 is a data-efficient version of TAU22, designed with reduced training data to simulate low-resource learning scenarios. The training subsets in TAU24 include approximately 5\%, 10\%, 25\%, 50\%, and 100\% of the training split from TAU22. In our experiments, we simply adopt the 25\% split for evaluation, consistent with the official setting of the DCASE2025 Challenge. To make consistent comparison with the DCASE solutions, the official train-test split setting is followed in all experiments.


\subsection{Backbones}
To validate the effectiveness of proposed City2Scene, we implement three lightweight Convolutional Neural Networks (CNNs) the as student models, and two Transformers as the teacher models in the ASC system. The CNN models, CP-ResNet \cite{koutini2021receptive}, BC-ResNet \cite{kim21l_interspeech}, and TF-SepNet \cite{cai2024tf}, were top-ranked solutions in recent DCASE Challenges with low-complexity design. Each model contains fewer than 128K parameters and requires less than 30 million multiply-accumulate operations (MACs) of computational overhead. CP-ResNet controls the size of receptive field to improve the generalization of models. BC-ResNet combines 1D temporal convolution with 2D convolution to capture representation of audio features with less computation. TF-SepNet separates the feature processing along the time and frequency dimensions to capture more time-frequency features in audio signals.

Unlike efficient CNN models, the Transformer models, PaSST \cite{koutini22_interspeech} and BEATs \cite{pmlr-v202-chen23ag, Cai2024workshop}, consume a large number of computational resources, making them primarily adopted as high-performing teacher models for knowledge distillation in recent submissions to DCASE Challenge \cite{Schmid2023workshop,Cai2024,han2024data}. Both PaSST and BEATs are pre-trained on AudioSet \cite{gemmeke2017audio}, which is a large-scale audio dataset annotated with various event labels. PaSST is a supervised Transformer model tailored for audio tasks, which uses a ``patchout" mechanism on spectrogram patches to capture long-range dependencies in acoustic features. BEATs, on the other hand, is a self-supervised pre-trained audio Transformer that leverages a bidirectional encoding mechanism to capture high-level acoustic features in data. In our experiments, we directly use the pre-trained checkpoints available in original works \cite{koutini22_interspeech, pmlr-v202-chen23ag}.

\subsection{Training Configuration}
We adopt different configurations of audio pre-processing, data augmentation and training strategy, tailored to each backbone model, mostly aligning with the intended design in the original works. City2Scene requires separating the feature extractor and classifier components of the backbone model. We follow the classifier settings from the original works and treat all previous layers as the feature extractor. Throughout all three stages of City2Scene, we mostly maintain a consistent configuration for each backbone model. The temperature factor $\tau$ in Equation (\ref{eq:city2scene}) is fixed to 2 as in \cite{Schmid2023workshop}, while the balance coefficient $\lambda$ in Equation (\ref{eq:loss}) is optimized as the hyperparameter to maximize the performance using the validation set. All experimental results presented are averaged over three independent runs with different seeds. Detailed hyperparameter settings are provided in the Section \ref{sec:appendix} of Appendix.

\begin{table}
    \centering
    \caption{Accuracy (\%) of Transformer-based city teachers in Stage 1 and Stage 2. \textbf{Scene:} scene classification. \textbf{City:} city classification. \textbf{Scene(City):} scene classification using frozen city features. \textbf{Ensemble:} ensemble model of PaSST and BEATs.}
    \begin{tabular}{l|l|c|c c}
    \multicolumn{2}{c|}{ }& Ref.& Stage 1& Stage 2\\
    \toprule
    & \textit{Model}& Scene& City& Scene(City)\\
    \midrule
    \multirow{4}{*}{\rotatebox{90}{TAU20}}& PaSST& 75.2$_{\pm 0.3}$& 39.2$_{\pm 0.3}$& 72.2$_{\pm 0.0}$\\
    & BEATs& 78.3$_{\pm 0.2}$& 48.2$_{\pm 0.1}$& 65.8$_{\pm 0.3}$\\
    \cmidrule{2-5}
    & Ensemble& 80.6$_{\pm 0.2}$& 50.0$_{\pm 0.2}$& 73.3$_{\pm 0.1}$\\
    \midrule
    \multirow{4}{*}{\rotatebox{90}{TAU22}}& PaSST& 60.2$_{\pm 0.1}$& 32.9$_{\pm 0.1}$& 54.8$_{\pm 0.0}$\\
    & BEATs& 61.7$_{\pm 0.2}$& 37.8$_{\pm 0.3}$& 51.1$_{\pm 0.1}$\\
    \cmidrule{2-5}
    & Ensemble& 64.9$_{\pm 0.1}$& 40.1$_{\pm 0.2}$& 58.0$_{\pm 0.0}$\\
    \midrule
    \multirow{4}{*}{\rotatebox{90}{TAU24}}& PaSST& 55.3$_{\pm 0.2}$& 25.5$_{\pm 0.2}$& 52.5$_{\pm 0.1}$\\
    & BEATs& 59.2$_{\pm 0.1}$& 32.3$_{\pm 0.2}$& 49.0$_{\pm 0.0}$\\
    \cmidrule{2-5}
    & Ensemble& 61.2$_{\pm 0.1}$& 33.2$_{\pm 0.2}$& 55.3$_{\pm 0.0}$\\
    \bottomrule
    \end{tabular}
    \label{tab:teacher_result}
\end{table}

\section{Results}
\label{sec:results}
This section elaborates the experimental results. Tables \ref{tab:teacher_result} and \ref{tab:student_result} present the performance of proposed City2Scene using two Transformer teacher models (PaSST and BEATs) and three student models (CP-ResNet, BC-ResNet and TF-SepNet) across three stages on the TAU20, TAU22 and TAU24 datasets. Table \ref{tab:acc_diff_teachers} shows the results of distilling knowledge from different teachers, including the city teacher, the fusion teacher and the weighted teacher. Table \ref{tab:gamma_attn_result} presents values of the attention weights $\gamma$ after training. Figures \ref{fig:tsne_confusion} and \ref{fig:acc_diff_methods} compare the performance of City2Scene with existing methods, muti-task learning and feature disentanglement. Figures \ref{fig:lambda} show the influence of the hyperparameter $\lambda$ to system.

\subsection{City Classification}
The Stage 1 column of Table \ref{tab:teacher_result} presents the city classification performance of Transformer-based teacher models PaSST and BEATs, along with their ensemble, across the TAU20, TAU22, and TAU24 datasets. It can be seen that city classification generally yields lower accuracy compared to scene classification across all datasets. For example, PaSST achieves 75.2\% accuracy in scene classification but only 39.2\% in city classification on TAU20. Similarly, BEATs shows a strong 78.3\% on scenes but drops to 48.2\% for cities. The discrepancy is likely due to the complexity of city classification task. Some cities may share similar environmental and cultural soundscapes, which requires the model to identify more discriminative cues in the acoustic features. With the advantages of self-supervised pre-training, BEATs consistently outperforms PaSST on city classification across all datasets. The ensemble of BEATs and PaSST further boosts city classification accuracy, reaching 50.0\% on TAU20, 40.1\% on TAU22, and 33.2\% on TAU24.

\subsection{Scene Classification using City Features}
\label{ssec:city_features}
To evaluate whether city features learned from city classification models can benefit scene classification, we conduct experiments using frozen features extracted from city teachers. The results are presented in the Stage 2 column of Table \ref{tab:teacher_result}. Compared to city classification in Stage 1, city features achieve surprisingly competitive results in classifying scenes, approaching the baseline scene accuracy. For example, PaSST achieves 72.2\% with frozen city features—only 3.0\% below its scene classification performance. The ensemble of PaSST and BEATs further boosts performance to 73.3\%. A similar trend is observed on TAU22 and TAU24. These results confirm that city classification models capture rich acoustic representations that generalize well to scene classification, even under reduced data conditions.

\begin{table}
    \centering
    \caption{Accuracy (\%) of CNN-based scene students in Stage 3, distilling knowledge from different city teachers . \textbf{Ref.:} baseline without City2Scene as reference. Best performance for each student is highlighted in bold.}
    \begin{tabular}{l|l|c|c c c}
    \toprule
    & \textit{Model}& Ref.& PaSST& BEATs& Ensemble\\
    \midrule
    \multirow{3}{*}{\rotatebox{90}{TAU20}}& CP-ResNet& 69.5$_{\pm 0.7}$& 69.6$_{\pm 0.4}$& \textbf{70.0}$_{\pm 0.7}$& 70.0$_{\pm 0.4}$ \\
    & BC-ResNet& 66.5$_{\pm 0.4}$& 66.9$_{\pm 0.8}$& \textbf{67.1}$_{\pm 0.4}$& 66.9$_{\pm 0.6}$ \\
    & TF-SepNet& 70.3$_{\pm 0.6}$& 70.3$_{\pm 1.0}$& 70.5$_{\pm 0.8}$& \textbf{71.3}$_{\pm 0.9}$ \\
    \midrule
    \multirow{3}{*}{\rotatebox{90}{TAU22}}& CP-ResNet& 55.2$_{\pm 0.5}$& \textbf{56.1}$_{\pm 0.4}$& 55.9$_{\pm 0.3}$& 56.1$_{\pm 0.4}$ \\
    &BC-ResNet& 59.4$_{\pm 0.3}$& 59.9$_{\pm 0.3}$& \textbf{60.4}$_{\pm 0.5}$& 60.1$_{\pm 0.7}$\\
    & TF-SepNet& 61.1$_{\pm 0.1}$& 61.3$_{\pm 0.1}$& 61.9$_{\pm 0.5}$& \textbf{62.1}$_{\pm 0.5}$ \\
    \midrule
    \multirow{3}{*}{\rotatebox{90}{TAU24}}& CP-ResNet& 51.3$_{\pm 0.6}$& 51.8$_{\pm 1.1}$& 51.9$_{\pm 0.9}$& \textbf{52.0}$_{\pm 0.6}$ \\
    &BC-ResNet& 51.5$_{\pm 0.2}$& 52.1$_{\pm 0.0}$& 52.2$_{\pm 0.1}$& \textbf{52.7}$_{\pm 0.3}$\\
    & TF-SepNet& 55.3$_{\pm 0.2}$& 55.9$_{\pm 0.1}$& \textbf{56.8}$_{\pm 0.8}$& 56.0$_{\pm 0.5}$ \\
    \bottomrule
    \end{tabular}
    \label{tab:student_result}
\end{table}

\subsection{Scene Classification with City Teachers}
As shown in Table~\ref{tab:student_result}, this subsection reports the performance of CNN-based scene student models (CP-ResNet, BC-ResNet, and TF-SepNet) in Stage 3 of the City2Scene framework, where knowledge is distilled from different Transformer-based city teachers. Each student is trained by distilling city-specific knowledge from PaSST, BEATs, and their ensemble, and the results are compared to a baseline trained without City2Scene (Ref.). Across all settings, City2Scene consistently improves scene classification performance over the baseline, demonstrating the effectiveness of our solution. Notably, the best-performing city teacher varies across datasets and backbones. On TAU20, BEATs provides the strongest single-teacher supervision for CP-ResNet and BC-ResNet, while the ensemble achieves the highest accuracy for TF-SepNet. On TAU22, the ensemble or BEATs again leads to the best results for all students, with TF-SepNet reaching the best accuracy of 62.1\%. On TAU24, which has reduced training data, TF-SepNet still achieves a performance gain of 1.2\% when distilling from BEATs.

\begin{table}
    \centering
    \caption{Knowledge Fusion of City and Scene Teachers.}
    \begin{tabular}{l|l|c c c}
    \toprule
    &\textit{Model}& City& Mean& Attention\\
    \midrule
    \multirow{3}{*}{\rotatebox{90}{TAU20}}& CP-ResNet& 70.0$_{\pm 0.7}$& 70.8$_{\pm 0.6}$& \textbf{71.1}$_{\pm 0.9}$ \\
    &BC-ResNet& 67.1$_{\pm 0.4}$& 67.6$_{\pm 0.6}$& \textbf{68.4}$_{\pm 0.5}$ \\
    &TF-SepNet& 71.3$_{\pm 0.9}$& 71.9$_{\pm 0.7}$& \textbf{72.5}$_{\pm 0.6}$ \\
    \midrule
    \multirow{3}{*}{\rotatebox{90}{TAU22}}& CP-ResNet& 56.1$_{\pm 0.4}$& 56.8$_{\pm 0.5}$& \textbf{57.0}$_{\pm 0.5}$ \\
    &BC-ResNet& 60.4$_{\pm 0.5}$& 61.0$_{\pm 0.1}$& \textbf{61.9}$_{\pm 0.7}$ \\
    &TF-SepNet& 62.1$_{\pm 0.5}$& 62.2$_{\pm 0.3}$& \textbf{63.5}$_{\pm 0.2}$ \\
    \midrule
    \multirow{3}{*}{\rotatebox{90}{TAU24}}& CP-ResNet& 52.0$_{\pm 0.6}$& 52.4$_{\pm 0.6}$& \textbf{52.5}$_{\pm 0.3}$ \\
    &BC-ResNet& 52.7$_{\pm 0.3}$& 52.9$_{\pm 0.5}$& \textbf{53.3}$_{\pm 0.5}$ \\
    &TF-SepNet& 56.8$_{\pm 0.8}$& 57.1$_{\pm 0.7}$& \textbf{57.5}$_{\pm 0.9}$ \\
    \bottomrule
    \end{tabular}
    \label{tab:acc_diff_teachers}
\end{table}

\begin{table}
    \centering
    \caption{Values of the average attention weight $\gamma$ after training.}
    \begin{tabular}{l|c c c}
    \toprule
    \textit{Model}& TAU20& TAU22& TAU24\\
    \midrule
    CP-ResNet& 0.710&0.677 &0.706 \\
    BC-ResNet& 0.526&0.596 &0.587 \\
    TF-SepNet& 0.453& 0.607&0.489 \\
    \bottomrule
    \end{tabular}
    \label{tab:gamma_attn_result}
\end{table}

\subsection{Fusion of City and Scene Knowledge}
The results of knowledge fusion from city and scene teachers are shown in Table \ref{tab:acc_diff_teachers}. The three columns respectively represent the results of distilling from city teachers only, a mean fusion of city and scene logits, and a trainable attention-based fusion. The mean fusion strategy consistently outperforms the city-only teacher setup across all backbones and datasets. Notably, the attention-based fusion yields the best performance, suggesting that adaptively balancing the contributions of city and scene knowledge leads to more effective supervision. For example, on TAU22, BC-ResNet improves from 60.4\% (city teacher only) to 61.9\% with attention-based fusion, and TF-SepNet reaches a new best of 63.5\%, with a gain of 1.4\%. These results highlight the complementary nature of city and scene knowledge. The fusion knowledge enables City2Scene benefit not only from city-aware supervision but also from task-specific scene guidance.

To better understand how the model balances knowledge from city and scene teachers during distillation, we analyze the average values of $\gamma$ learned by the attention-based fusion module as shown in Table \ref{tab:gamma_attn_result}. $\gamma$ in Equation \ref{eq:attn} reflects the relative importance assigned to the city teacher logits in the fused representation. Across all datasets and backbones, $\gamma$ remains above 0.45, indicating that the model consistently assigns considerable weight to city-specific knowledge. For CP-ResNet, the attention weights are particularly high, demonstrating the significant contribution of the city teacher. For stronger backbones like TF-SepNet, the attention weights are substantial, highlighting that city-specific knowledge provide valuable complementary information to scene-focused representations. Overall, the attention module selectively fuses teacher knowledge in a model-aware and data-aware manner. The dynamic weighting strategy contributes to improved performance without requiring manual tuning.

\begin{figure*}
    \centering
    \includegraphics[width=\linewidth]{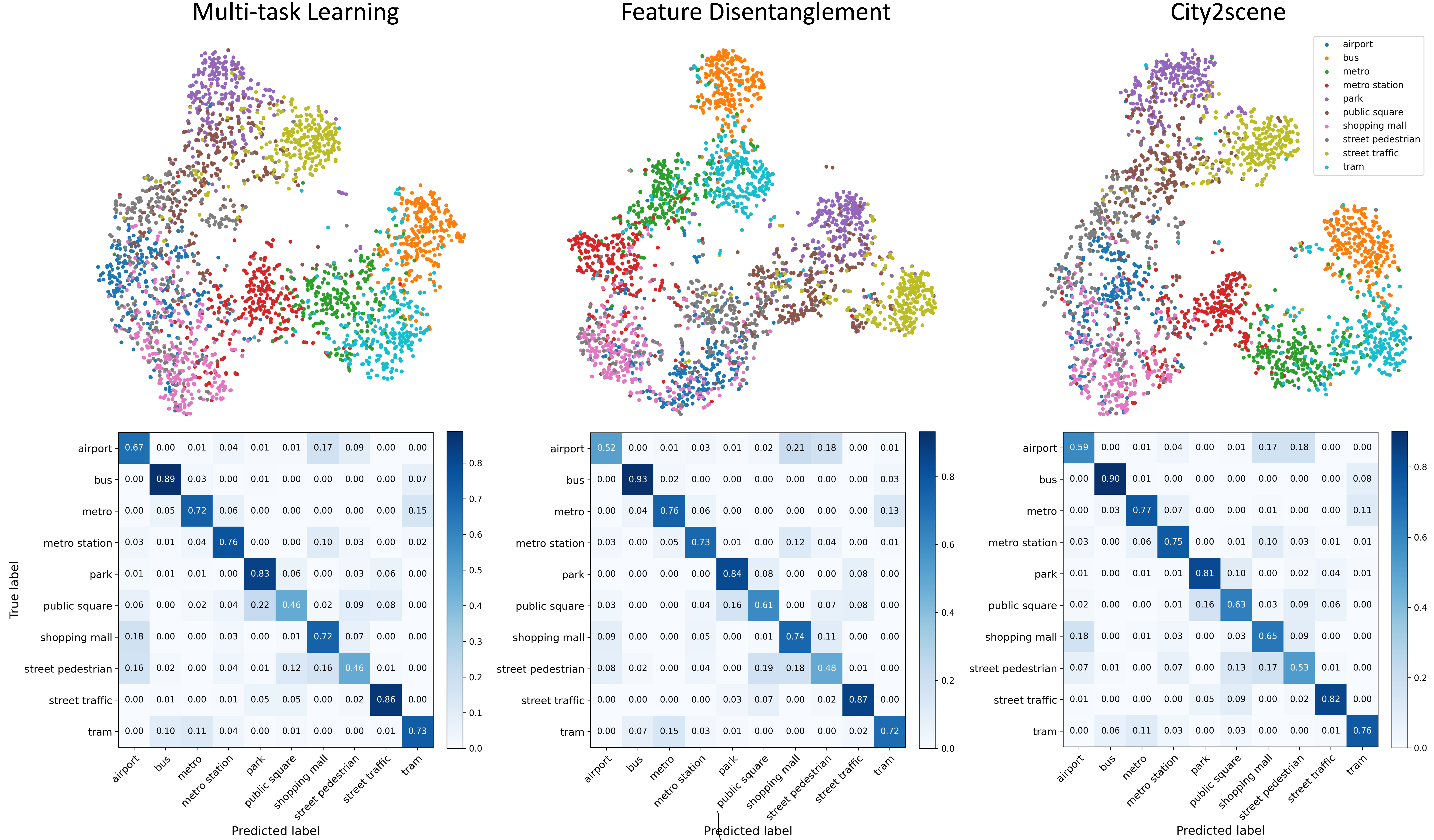}
    \caption{t-SNE visualization of scene classifier outputs of TF-SepNet and corresponding confusion matrices on TAU20 test set.}
    \label{fig:tsne_confusion}
\end{figure*}

\begin{figure}
    \centering
    \includegraphics[width=\linewidth]{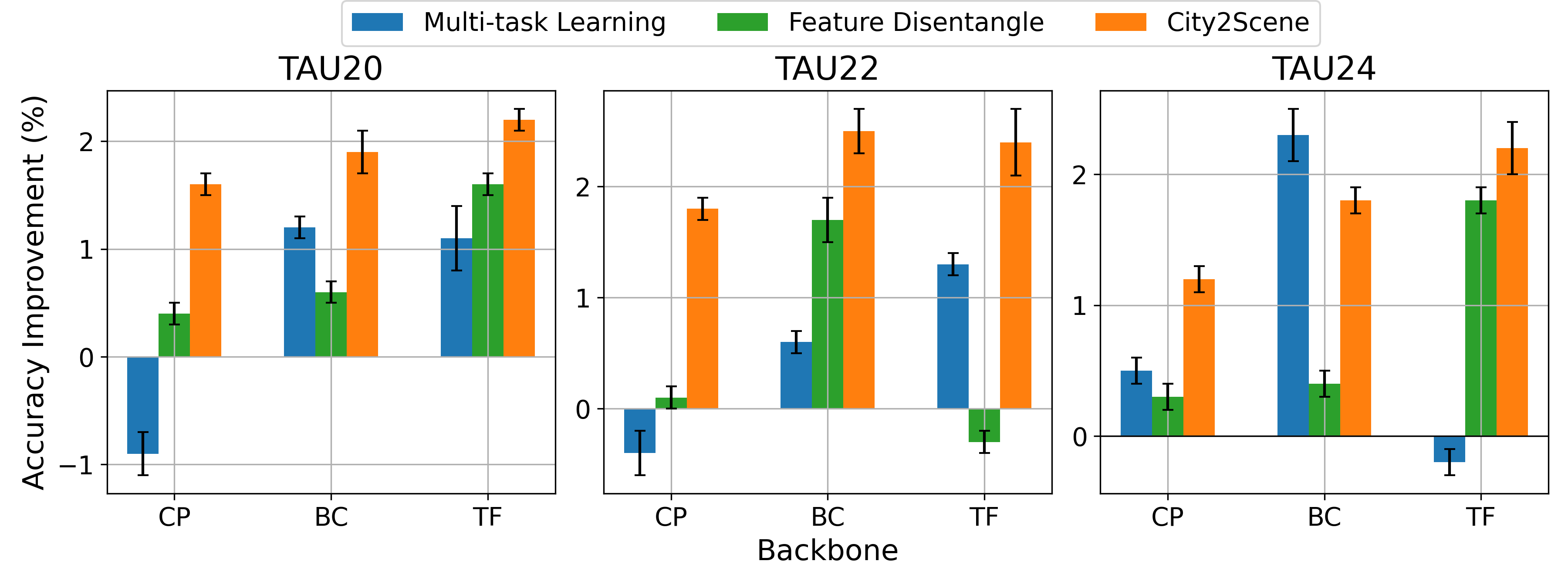}
    \caption{Accuracy improvements (\%) after applying different methods.}
    \label{fig:acc_diff_methods}
\end{figure}

\subsection{Comparison to Existing Methods}
To further validate the effectiveness of City2Scene, we compare it against two existing approaches described in Section \ref{ssec:related_methods}, multi-task learning and feature disentanglement. Figure \ref{fig:acc_diff_methods} shows the accuracy differences of each method relative to the baseline. Multi-task learning yields improvements in some settings (e.g., +2.3\% for BC-ResNet on TAU24), but is less stable and even leads to performance degradation in others (e.g., -0.9\% for CP-ResNet on TAU20). This instability likely stems from the conflict between the scene and city objectives during joint optimization, which can hinder the feature learning for the main task. Feature disentanglement shows more stable gains than multi-task learning on TAU20 and TAU24, but underperforms when applied to TF-SepNet on TAU22. The possible reason is that domain-invariant features encouraged by the gradient reversal mechanism may suppress useful scene-specific features, particularly when domain gaps are small or when model capacity is limited. In contrast, City2Scene consistently outperforms all other methods across datasets and backbones.

Figure \ref{fig:tsne_confusion} visualizes the learned feature distributions and corresponding confusion matrices to further illustrate the differences between methods beyond accuracy. Figure \ref{fig:tsne_confusion}(a), corresponding to MTL, shows substantial overlap between semantically similar scenes such as \textit{tram}, \textit{bus}, \textit{metro}, and \textit{metro station}. These classes are acoustically similar and often hard to distinguish, even for human listeners. Compared to Figure \ref{fig:tsne_confusion}(a) and \ref{fig:tsne_confusion}(b), Figure \ref{fig:tsne_confusion}(c) exhibits significantly clearer class boundaries. Notably, \textit{tram} forms a compact cluster, better separated from \textit{metro} and \textit{bus}. Similarly, \textit{metro station} samples are more centrally clustered and less entangled with other classes. The accompanying confusion matrices also provide a finer-grained view of inter-class confusion. City2Scene reduces confusion between overlapping classes, particularly within public transport-related scenes.

\subsection{Comparison to Top-ranked DCASE Solutions}
To assess the competitiveness of City2Scene, we compare it against top-ranked systems from the DCASE Challenge Task 1 over the past four years (2021–2024). Table~\ref{tab:dcase} summarizes the best reported accuracies from the official challenge leaderboards. City2Scene consistently outperforms the DCASE baseline systems by a large margin in all years. On the 2022 and 2023 dataset (same data), City2Scene-Attention reaches 63.6\%, surpassing the Top-1 systems. On the 2021 and 2024 dataset, City2Scene-Attention respectively achieves 73.0\% and 58.3\%, just below the best systems.

\begin{table}
    \centering
    \caption{Comparison to the top-ranked systems in recent DCASE Challenge task 1. Note: 2022 and 2023 used the same dataset.}
    \begin{tabular}{l|c c c c}
        \toprule
        \textit{System}& 2021& 2022& 2023& 2024\\
        \midrule
        DCASE Baseline& 47.7& 42.9& 42.9& 50.3\\
        Top-1 Accuracy& \textbf{76.1} \cite{Kim2021b}& 60.8 \cite{Lee2022}& 62.7 \cite{Schmid2023}& \textbf{59.1} \cite{Bing2024}\\
        Top-2 Accuracy& 72.9 \cite{Yang2021}& 59.7 \cite{Schmid2022}& 60.8 \cite{Tan2023a}& 58.3 \cite{Shao2024}\\
        Top-3 Accuracy& 72.1 \cite{Koutini2021}& 56.3 \cite{Xin2022}& 57.0 \cite{Cai2023a}& 58.1 \cite{Cai2024}\\
        \midrule
        City2Scene-City& 71.9& 62.4& 62.4& 57.1\\
        City2Scene-Mean& 72.5& 62.5& 62.5& 57.8\\
        City2Scene-Attention& 73.0& \textbf{63.6}& \textbf{63.6}& 58.3\\
        \bottomrule
    \end{tabular}
    \label{tab:dcase}
\end{table}

\begin{figure}
    \centering
    \includegraphics[width=\linewidth]{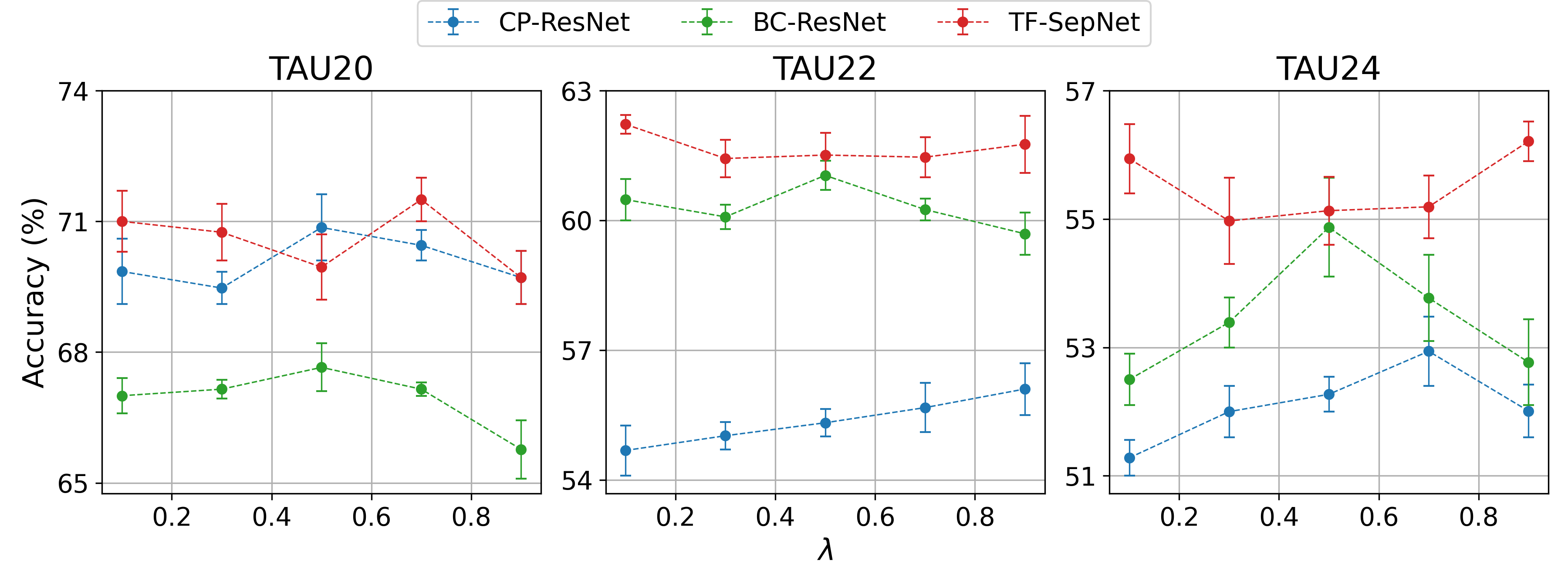}
    \caption{Impact of hyperparameter $\lambda$.}
    \label{fig:lambda}
\end{figure}

\subsection{Sensitivity to Hyperparameter}
The hyperparameter $\lambda$ in Equation \ref{eq:loss} balances the trade-off between label and distillation loss. The sensitivity of the models to varying $\lambda$ values (from 0.1 to 0.9) is presented in Figure \ref{fig:lambda}. Higher values of $\lambda$ emphasize the label loss, while lower values give more weight to the distillation loss. On TAU20, all models demonstrate relatively stable performance, with the best accuracies achieved near the middle range. Specifically, CP-ResNet and BC-ResNet both peaks at $\lambda=0.5$, while TF-SepNet sees its highest accuracy at $\lambda=0.7$. In contrast, TAU22 shows a gradual increase in performance for CP-ResNet as $\lambda$ increases. BC-ResNet performs best at $\lambda=0.5$ and TF-SepNet maintains a relatively flat trend with a slight rise at the start point. On the more challenging TAU24 dataset, BC-ResNet keeps the peak at $\lambda=0.5$, CP-ResNet and TF-SepNet respectively reach the best performance at $\lambda=0.7$ and $0.9$. Above results suggest that while moderate values of $\lambda$ (0.5-0.7) are generally effective, models trained under low-resource conditions (e.g., TAU24) may benefit more from a higher emphasis on the label loss. Furthermore, careful tuning of $\lambda$ is essential to maximize the benefit of City2Scene across different datasets.

\begin{figure}
    \centering
    \includegraphics[width=\linewidth]{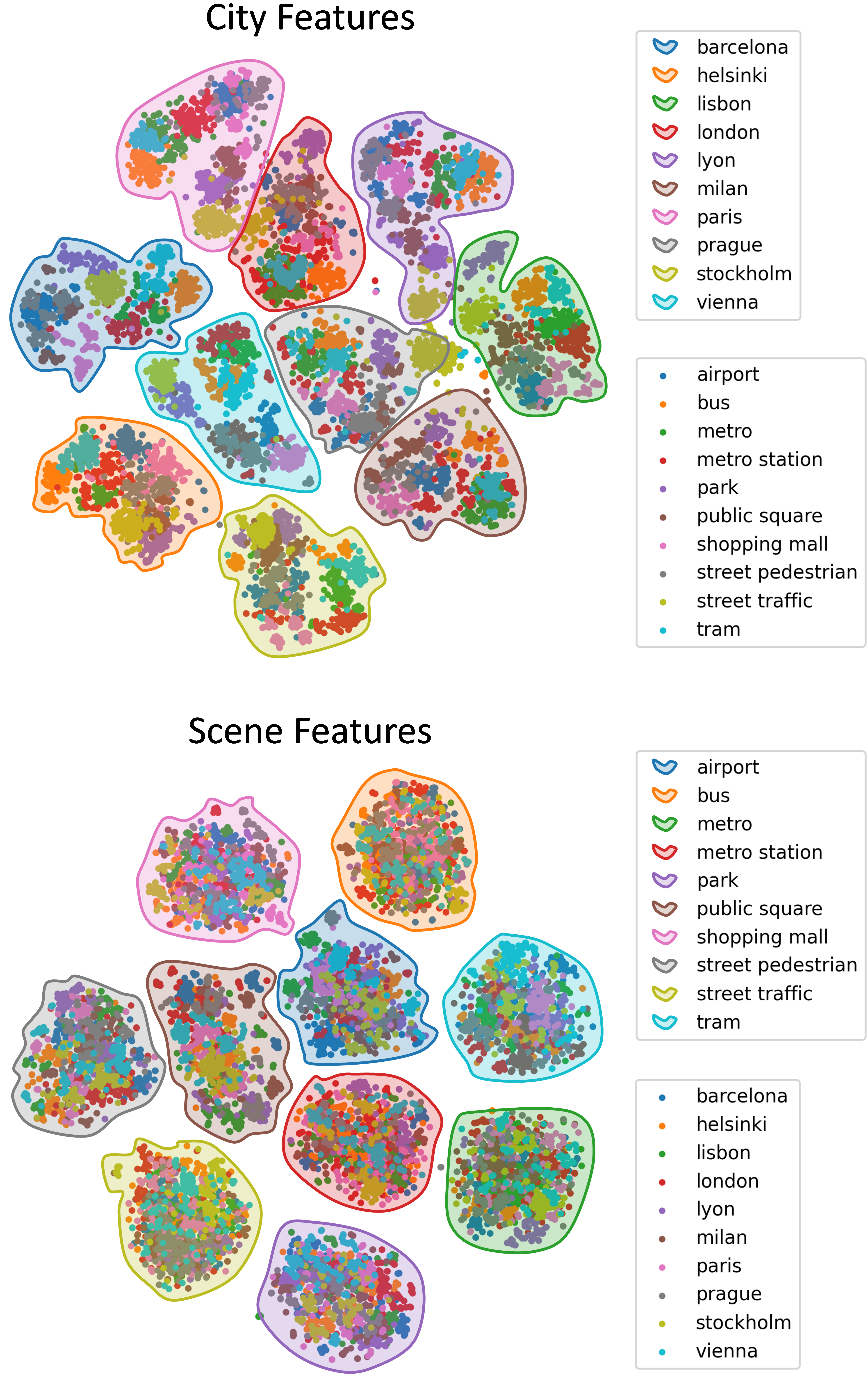}
    \caption{City features and scene features extracted by BEATs on TAU20.}
    \label{fig:tsne}
\end{figure}

\section{Discussion}
\label{sec:discussion}
This section offers deeper insights into the behavior of City2Scene based on the experimental findings. We discuss why city features provide benefits for ASC. In addition, we analyze when the improvements are most significant, and examine potential limitations.

The experiments in Section \ref{ssec:city_features} demonstrate that city features contain rich information that are discriminative for classifying scenes. The benefits of city features are further illustrated in Figure \ref{fig:tsne}, which visualizes the city and scene features extracted by BEATs on TAU20 training set. Both the top and bottom plots show samples cluster distinctly based on corresponding task labels, indicating that the training effectively captures task-specific acoustic patterns. Notably, in the top figure, features learned from city training also result in scene-based clustering, suggesting that city classification encodes acoustic information that is relevant and transferable to the scene classification task. In contrast, the bottom figure shows that features learned from scene training are more dispersed with respect to city labels. We also quantity this observation through estimating the mutual information (MI) between features and labels. Following the approach in prior work \cite{wang2021revisiting}, we adopt the test accuracy of an auxiliary classifier to estimate MI between feature representations $e$ and label $y$, using the formulation $I(e, y) = H(y) - \mathbb{E}_{p(e,y)} [\log p(y|e)]$, where the conditional distribution $p(y|e)$ is approximated using a classifier $q_\phi(y|e)$ trained on top of the feature embeddings. As shown in Table \ref{tab:mi}, the city features $f_c(x)$ show an MI of 0.658 with scene labels. Conversely, scene features $f_s(x)$ have lower MI of 0.306 with city labels. The phenomenon of asymmetry reveals the correlation of semantic information between cities and scenes within the learned representations. The relatively high MI between city features and scene labels suggests that city-trained models capture acoustic patterns that are not only useful for city discrimination but also highly informative for classifying scenes. Since cities often exhibit unique environmental characteristics such as transportation systems, infrastructure acoustics and population density, models trained to recognize cities may also capture high-level features that generalize well to scenes. In contrast, the lower MI from scene features to city labels implies that standard ASC models tend to focus narrowly on task-relevant features while discarding broader contextual or geographic information. As a result, scene features and city features may partially overlap, but city features capture a larger scale of acoustic information, some of which is useful for scene classification yet not preserved in scene-specific models. These findings support our core hypothesis: city features serve as a rich auxiliary source of knowledge. By distilling city-specific knowledge into ASC models, City2Scene leverages the underutilized information embedded in city features, providing a new way to improve acoustic scene classification.

\begin{table}
    \centering
    \caption{Estimates of mutual information between features and labels. $f_c(x)$: city features; $f_s(x)$: scene features; $y_c$: city labels; $y_c$: scene labels. Results of BEATs on TAU20 test set are reported.}
    \begin{tabular}{l|c c}
        \toprule
        \textit{Features}& $I(e,y_c)$& $I(e,y_s)$\\
        \midrule
        $e=f_c(x)$& 0.482& 0.658\\
        $e=f_s(x)$& 0.306& 0.783\\
        \bottomrule
    \end{tabular}
    \label{tab:mi}
\end{table}

The contribution of City2Scene is further shown in Table \ref{tab:citywise}, which presents city-wise ASC performance differences between baseline and City2Scene with TF-SepNet on TAU20. The accuracy in most cities has improved, with an average gain of 3.0\%. Cities such as \textit{Stockholm} (+6.3\%), \textit{Milan} (+4.8\%), and \textit{Helsinki} (+4.5\%) show significant improvements, when City2Scene is applied. It is possible that these cities have distinctive acoustic features strongly correlated with corresponding scene environments, making the city knowledge particularly beneficial. Interestingly, \textit{Prague} and \textit{Vienna} show slight performance drops of 0.4\%, which may occur when city-specific acoustic features are weak or inconsistent across scenes, potentially introducing noise during distillation. Nevertheless, the overall trend in the city-wise analysis highlights that City2Scene enhances the ASC performance by improving generalization across different cities.

\section{Conclusion}
\label{sec:conclusion}
In this paper, we proposed City2Scene, a novel framework for improving acoustic scene classification (ASC) by leveraging city features. Experimental results demonstrate the city features include valuable information for classifying scenes, and show the effectiveness of City2Scene across multiple backbone architectures and datasets. Furthermore, the knowledge fusion from city and scene teachers outperform the single city teacher, highlighting the importance of diverse task teachers for knowledge distillation in ASC. We also find that the improvements introduced by City2Scene are closely related to enhanced generalization across cities. Future work will investigate more methods to integrate city information into the ASC task. In summary, exploring potential relations between city features and acoustic scenes remains a promising direction for advancing ASC research.

\begin{table}
    \centering
    \caption{City-wise scene accuracy (\%) of TF-SepNet on TAU20.}
    \begin{tabular}{l|c c|l}
        \toprule
        \textit{City}& Scene& City2Scene& Diff.\\
        \midrule
        Barcelona& 76.0& 76.8& $+$0.8\\
        Helsinki& 58.6& 63.1& $+$4.5\\
        Lisbon& 74.5& 78.1& $+$3.7\\
        London& 73.7& 77.0& $+$3.3\\
        Lyon& 75.9& 79.3& $+$3.4\\
        Milan& 66.7& 71.5& $+$4.8\\
        Paris& 63.3& 67.0& $+$3.7\\
        Prague& 71.5& 71.1& $-$0.4\\
        Stockholm& 68.9& 75.2& $+$6.3\\
        Vienna& 66.3& 65.9& $-$0.4\\
        \midrule
        Average& 69.6& 72.5& $+$3.0\\
        \bottomrule
    \end{tabular}
    \label{tab:citywise}
\end{table}


\bibliographystyle{IEEEtran}
\bibliography{references}

\begin{thebibliography}{10}
\providecommand{\url}[1]{#1}
\csname url@samestyle\endcsname
\providecommand{\newblock}{\relax}
\providecommand{\bibinfo}[2]{#2}
\providecommand{\BIBentrySTDinterwordspacing}{\spaceskip=0pt\relax}
\providecommand{\BIBentryALTinterwordstretchfactor}{4}
\providecommand{\BIBentryALTinterwordspacing}{\spaceskip=\fontdimen2\font plus
\BIBentryALTinterwordstretchfactor\fontdimen3\font minus \fontdimen4\font\relax}
\providecommand{\BIBforeignlanguage}[2]{{%
\expandafter\ifx\csname l@#1\endcsname\relax
\typeout{** WARNING: IEEEtran.bst: No hyphenation pattern has been}%
\typeout{** loaded for the language `#1'. Using the pattern for}%
\typeout{** the default language instead.}%
\else
\language=\csname l@#1\endcsname
\fi
#2}}
\providecommand{\BIBdecl}{\relax}
\BIBdecl

\bibitem{barchiesi2015acoustic}
D.~Barchiesi, D.~Giannoulis, D.~Stowell, and M.~D. Plumbley, ``Acoustic scene classification: Classifying environments from the sounds they produce,'' \emph{IEEE Signal Processing Magazine}, vol.~32, no.~3, pp. 16--34, 2015.

\bibitem{Mesaros2018}
A.~Mesaros, T.~Heittola, and T.~Virtanen, ``A multi-device dataset for urban acoustic scene classification,'' in \emph{Proceedings of the Detection and Classification of Acoustic Scenes and Events 2018 Workshop (DCASE2018)}, November 2018, pp. 9--13.

\bibitem{Heittola2020}
T.~Heittola, A.~Mesaros, and T.~Virtanen, ``Acoustic scene classification in {DCASE} 2020 challenge: Generalization across devices and low complexity solutions,'' in \emph{Proceedings of the Detection and Classification of Acoustic Scenes and Events 2020 Workshop (DCASE2020)}, 2020, pp. 56--60.

\bibitem{bai2024description}
J.~Bai, M.~Wang, H.~Liu, H.~Yin, Y.~Jia, S.~Huang, Y.~Du, D.~Zhang, M.~D. Plumbley, D.~Shi \emph{et~al.}, ``Description on {IEEE} {ICME} 2024 grand challenge: Semi-supervised acoustic scene classification under domain shift,'' \emph{arXiv preprint arXiv:2402.02694}, 2024.

\bibitem{hu2021two}
H.~Hu, C.-H.~H. Yang, X.~Xia, X.~Bai, X.~Tang, Y.~Wang, S.~Niu, L.~Chai, J.~Li, H.~Zhu \emph{et~al.}, ``A two-stage approach to device-robust acoustic scene classification,'' in \emph{2021 IEEE International Conference on Acoustics, Speech and Signal Processing (ICASSP)}.\hskip 1em plus 0.5em minus 0.4em\relax IEEE, 2021, pp. 845--849.

\bibitem{singh2021prototypical}
S.~Singh, H.~L. Bear, and E.~Benetos, ``Prototypical networks for domain adaptation in acoustic scene classification,'' in \emph{2021 IEEE International Conference on Acoustics, Speech and Signal Processing (ICASSP)}.\hskip 1em plus 0.5em minus 0.4em\relax IEEE, 2021, pp. 346--350.

\bibitem{liu2021cross}
Y.~Liu, A.~Neophytou, S.~Sengupta, and E.~Sommerlade, ``Cross-modal spectrum transformation network for acoustic scene classification,'' in \emph{2021 IEEE International Conference on Acoustics, Speech and Signal Processing (ICASSP)}.\hskip 1em plus 0.5em minus 0.4em\relax IEEE, 2021, pp. 830--834.

\bibitem{yan2024semi}
Y.~Yan, W.~Liu, Y.~Chai, and Y.~Ren, ``Semi-supervised acoustic scene classification under domain shift with mixmatch and information bottleneck optimization,'' in \emph{2024 IEEE International Conference on Multimedia and Expo Workshops (ICMEW)}.\hskip 1em plus 0.5em minus 0.4em\relax IEEE, 2024, pp. 1--4.

\bibitem{bear2019city}
H.~L. Bear, T.~Heittola, A.~Mesaros, E.~Benetos, and T.~Virtanen, ``City classification from multiple real-world sound scenes,'' in \emph{2019 IEEE Workshop on Applications of Signal Processing to Audio and Acoustics (WASPAA)}.\hskip 1em plus 0.5em minus 0.4em\relax IEEE, 2019, pp. 11--15.

\bibitem{tan2024acoustic}
Y.~Tan, H.~Ai, S.~Li, and M.~D. Plumbley, ``Acoustic scene classification across cities and devices via feature disentanglement,'' \emph{IEEE/ACM Transactions on Audio, Speech, and Language Processing}, vol.~32, pp. 1286--1297, 2024.

\bibitem{Schmid2023workshop}
F.~Schmid, T.~Morocutti, S.~Masoudian, K.~Koutini, and G.~Widmer, ``Distilling the knowledge of transformers and {CNNs} with {CP}-mobile,'' in \emph{Proceedings of the Detection and Classification of Acoustic Scenes and Events 2023 Workshop (DCASE2023)}, 2023, pp. 161--165.

\bibitem{Martin-Morato2022}
I.~Mart\'{i}n-Morat\'{o}, F.~Paissan, A.~Ancilotto, T.~Heittola, A.~Mesaros, E.~Farella, A.~Brutti, and T.~Virtanen, ``Low-complexity acoustic scene classification in dcase 2022 challenge,'' in \emph{Proceedings of the 7th Detection and Classification of Acoustic Scenes and Events 2022 Workshop (DCASE2022)}, November 2022.

\bibitem{Schmid2024a}
F.~Schmid, P.~Primus, T.~Heittola, A.~Mesaros, I.~Martín-Morató, K.~Koutini, and G.~Widmer, ``Data-efficient low-complexity acoustic scene classification in the dcase 2024 challenge,'' in \emph{Proceedings of the Detection and Classification of Acoustic Scenes and Events 2024 Workshop (DCASE2024)}, October 2024, pp. 136--140.

\bibitem{Caruana1998}
R.~Caruana, \emph{Multitask Learning}.\hskip 1em plus 0.5em minus 0.4em\relax Springer US, 1998, pp. 95--133.

\bibitem{ganin2016domain}
Y.~Ganin, E.~Ustinova, H.~Ajakan, P.~Germain, H.~Larochelle, F.~Laviolette, M.~March, and V.~Lempitsky, ``Domain-adversarial training of neural networks,'' \emph{Journal of machine learning research}, vol.~17, no.~59, pp. 1--35, 2016.

\bibitem{Martin2021}
I.~Martin, T.~Heittola, A.~Mesaros, and T.~Virtanen, ``Low-complexity acoustic scene classification for multi-device audio: Analysis of dcase 2021 challenge systems,'' in \emph{Proceedings of the 6th Detection and Classification of Acoustic Scenes and Events 2021 Workshop (DCASE2021)}, November 2021, pp. 85--89.

\bibitem{singh22_interspeech}
A.~Singh and M.~D. Plumbley, ``A passive similarity based cnn filter pruning for efficient acoustic scene classification,'' in \emph{Proceedings of the Conference of the International Speech Communication Association (INTERSPEECH)}.\hskip 1em plus 0.5em minus 0.4em\relax ISCA, 2022, pp. 2433--2437.

\bibitem{koutini2019receptive}
K.~Koutini, H.~Eghbal-Zadeh, M.~Dorfer, and G.~Widmer, ``The receptive field as a regularizer in deep convolutional neural networks for acoustic scene classification,'' in \emph{2019 27th European signal processing conference (EUSIPCO)}.\hskip 1em plus 0.5em minus 0.4em\relax IEEE, 2019, pp. 1--5.

\bibitem{kim21l_interspeech}
B.~Kim, S.~Chang, J.~Lee, and D.~Sung, ``{Broadcasted residual learning for efficient keyword spotting},'' in \emph{Proceedings of the Conference of the International Speech Communication Association (INTERSPEECH)}.\hskip 1em plus 0.5em minus 0.4em\relax ISCA, 2021, pp. 4538--4542.

\bibitem{cai2024tf}
Y.~Cai, P.~Zhang, and S.~Li, ``{TF-SepNet}: An efficient {1D} kernel design in {CNNs} for low-complexity acoustic scene classification,'' in \emph{2024 IEEE International Conference on Acoustics, Speech and Signal Processing (ICASSP)}.\hskip 1em plus 0.5em minus 0.4em\relax IEEE, 2024, pp. 821--825.

\bibitem{hinton2015distilling}
G.~Hinton, O.~Vinyals, and J.~Dean, ``Distilling the knowledge in a neural network,'' \emph{arXiv preprint arXiv:1503.02531}, 2015.

\bibitem{schmid2022knowledge}
F.~Schmid, S.~Masoudian, K.~Koutini, and G.~Widmer, ``Knowledge distillation from transformers for low-complexity acoustic scene classification,'' in \emph{Proceedings of the Detection and Classification of Acoustic Scenes and Events 2022 Workshop (DCASE2022)}, 2022.

\bibitem{han2024data}
B.~Han, W.~Huang, Z.~Chen, A.~Jiang, P.~Fan, C.~Lu, Z.~Lv, J.~Liu, W.-Q. Zhang, and Y.~Qian, ``Data-efficient low-complexity acoustic scene classification via distilling and progressive pruning,'' \emph{arXiv preprint arXiv:2410.20775}, 2024.

\bibitem{koutini2021receptive}
K.~Koutini, H.~Eghbal-zadeh, and G.~Widmer, ``Receptive field regularization techniques for audio classification and tagging with deep convolutional neural networks,'' \emph{IEEE/ACM Transactions on Audio, Speech, and Language Processing}, vol.~29, pp. 1987--2000, 2021.

\bibitem{koutini22_interspeech}
K.~Koutini, J.~Schlüter, H.~Eghbal-zadeh, and G.~Widmer, ``Efficient training of audio transformers with patchout,'' in \emph{Proceedings of the Conference of the International Speech Communication Association (INTERSPEECH)}.\hskip 1em plus 0.5em minus 0.4em\relax ISCA, 2022, pp. 2753--2757.

\bibitem{pmlr-v202-chen23ag}
S.~Chen, Y.~Wu, C.~Wang, S.~Liu, D.~Tompkins, Z.~Chen, W.~Che, X.~Yu, and F.~Wei, ``{BEAT}s: Audio pre-training with acoustic tokenizers,'' in \emph{Proceedings of the 40th International Conference on Machine Learning (ICML)}.\hskip 1em plus 0.5em minus 0.4em\relax PMLR, 2023, pp. 5178--5193.

\bibitem{Cai2024workshop}
Y.~Cai, S.~Li, and X.~Shao, ``Leveraging self-supervised audio representations for data-efficient acoustic scene classification,'' in \emph{Proceedings of the Detection and Classification of Acoustic Scenes and Events 2024 Workshop (DCASE2024)}, October 2024, pp. 21--25.

\bibitem{Cai2024}
Y.~Cai, M.~Lin, S.~Li, and X.~Shao, ``{DCASE2024} task1 submission: Data-efficient acoustic scene classification with self-supervised teachers,'' DCASE2024 Challenge, Tech. Rep., May 2024.

\bibitem{gemmeke2017audio}
J.~F. Gemmeke, D.~P. Ellis, D.~Freedman, A.~Jansen, W.~Lawrence, R.~C. Moore, M.~Plakal, and M.~Ritter, ``Audio set: An ontology and human-labeled dataset for audio events,'' in \emph{2017 IEEE international conference on acoustics, speech and signal processing (ICASSP)}.\hskip 1em plus 0.5em minus 0.4em\relax IEEE, 2017, pp. 776--780.

\bibitem{Kim2021b}
B.~Kim, S.~Yang, J.~Kim, and S.~Chang, ``{QTI} submission to {DCASE} 2021: Residual normalization for device-imbalanced acoustic scene classification with efficient design,'' DCASE2021 Challenge, Tech. Rep., June 2021.

\bibitem{Lee2022}
J.-H. Lee, J.-H. Choi, P.~M. Byun, and J.-H. Chang, ``Hyu submission for the {DCASE} 2022: Efficient fine-tuning method using device-aware data-random-drop for device-imbalanced acoustic scene classification,'' DCASE2022 Challenge, Tech. Rep., June 2022.

\bibitem{Schmid2023}
F.~Schmid, T.~Morocutti, S.~Masoudian, K.~Koutini, and G.~Widmer, ``{CP-JKU} submission to dcase23: Efficient acoustic scene classification with cp-mobile,'' DCASE2023 Challenge, Tech. Rep., May 2023.

\bibitem{Bing2024}
H.~Bing, H.~Wen, C.~Zhengyang, J.~Anbai, C.~Xie, F.~Pingyi, L.~Cheng, L.~Zhiqiang, L.~Jia, Z.~Wei-Qiang, and Q.~Yanmin, ``Data-efficient acoustic scene classification via ensemble teachers distillation and pruning,'' DCASE2024 Challenge, Tech. Rep., May 2024.

\bibitem{Yang2021}
C.-H.~H. Yang, H.~Hu, S.~M. Siniscalchi, Q.~Wang, W.~Yuyang, X.~Xia, Y.~Zhao, Y.~Wu, Y.~Wang, J.~Du, and C.-H. Lee, ``A lottery ticket hypothesis framework for low-complexity device-robust neural acoustic scene classification,'' DCASE2021 Challenge, Tech. Rep., June 2021.

\bibitem{Schmid2022}
F.~Schmid, S.~Masoudian, K.~Koutini, and G.~Widmer, ``{CP-JKU} submission to {DCASE22}: Distilling knowledge for low-complexity convolutional neural networks from a patchout audio transformer,'' Detection and Classification of Acoustic Scenes and Events (DCASE) Challenge, Tech. Rep., 2022.

\bibitem{Tan2023a}
J.~Tan and Y.~Li, ``Low-complexity acoustic scene classification using blueprint separable convolution and knowledge distillation,'' DCASE2023 Challenge, Tech. Rep., May 2023.

\bibitem{Shao2024}
Y.-F. Shao, P.~Jiang, and W.~Li, ``Low-complexity acoustic scene classification with limited training data,'' DCASE2024 Challenge, Tech. Rep., May 2024.

\bibitem{Koutini2021}
K.~Koutini, S.~Jan, and G.~Widmer, ``Cpjku submission to dcase21: Cross-device audio scene classification with wide sparse frequency-damped {CNNs},'' DCASE2021 Challenge, Tech. Rep., June 2021.

\bibitem{Xin2022}
Y.~Xin, Y.~Zou, F.~Cui, and Y.~Wang, ``Low-complexity acoustic scene classification with mismatch-devices using separable convolutions and coordinate attention,'' DCASE2022 Challenge, Tech. Rep., June 2022.

\bibitem{Cai2023a}
Y.~Cai, M.~Lin, C.~Zhu, S.~Li, and X.~Shao, ``Dcase2023 task1 submission: Device simulation and time-frequency separable convolution for acoustic scene classification,'' DCASE2023 Challenge, Tech. Rep., May 2023.

\bibitem{wang2021revisiting}
Y.~Wang, Z.~Ni, S.~Song, L.~Yang, and G.~Huang, ``Revisiting locally supervised learning: an alternative to end-to-end training,'' in \emph{International Conference on Learning Representations (ICLR)}, 2021.

\bibitem{zhang2018mixup}
H.~Zhang, M.~Cisse, Y.~N. Dauphin, and D.~Lopez-Paz, ``mixup: Beyond empirical risk minimization,'' in \emph{Proceedings of the International Conference on Learning Representations (ICLR)}, 2018.

\bibitem{Park2019}
D.~S. Park, W.~Chan, Y.~Zhang, C.-C. Chiu, B.~Zoph, E.~D. Cubuk, and Q.~V. Le, ``Specaugment: A simple data augmentation method for automatic speech recognition,'' in \emph{Proceedings of the Conference of the International Speech Communication Association (INTERSPEECH)}.\hskip 1em plus 0.5em minus 0.4em\relax ISCA, 2019.

\bibitem{morocutti2023device}
T.~Morocutti, F.~Schmid, K.~Koutini, and G.~Widmer, ``Device-robust acoustic scene classification via impulse response augmentation,'' in \emph{2023 31st European Signal Processing Conference (EUSIPCO)}.\hskip 1em plus 0.5em minus 0.4em\relax IEEE, 2023, pp. 176--180.

\bibitem{loshchilov2017sgdr}
I.~Loshchilov and F.~Hutter, ``{SGDR}: Stochastic gradient descent with warm restarts,'' in \emph{Proceedings of the International Conference on Learning Representations (ICLR)}, 2017.

\end{thebibliography}

\section*{Appendix}
\label{sec:appendix}
Table \ref{tab:appendix} shows the hyperparameter settings in experiments.

\begin{table*}
    \centering
    \caption{Hyperparameters of audio preprocessing, data augmentation and training configuration for various backbone models.}
    \begin{tabular}{c|l|c|c|c|c|c}
        \toprule
        \multicolumn{2}{c|}{\textit{Hyperparameter}}& BC-ResNet& TF-SepNet& CP-ResNet& PaSST& BEATs\\
        \midrule
        \multirow{4}{*}{\rotatebox{0}{Preprocessing}}& Sampling rate& \multicolumn{4}{c|}{32kHz}& 16kHz\\
        & Window length& \multicolumn{3}{c|}{96ms}& \multicolumn{2}{c}{25ms}\\
        & Hop length& \multicolumn{2}{c|}{16ms}& 23ms& \multicolumn{2}{c}{10ms}\\
        & Mel bins& 256& 512& 256& \multicolumn{2}{c}{128}\\
        \midrule
        \multirow{4}{*}{\rotatebox{0}{Augmentation}}& Mixup ($\gamma$) \cite{zhang2018mixup}& \multicolumn{3}{c|}{0.3}& -& 0.3\\
        & SpecAug ($r, p$) \cite{Park2019}& \multicolumn{3}{c|}{-}& -& 0.2, 1.0\\
        & Freq-MixStyle ($\gamma,p$) \cite{Schmid2022}& \multicolumn{3}{c|}{0.4, 0.8}& \multicolumn{2}{c}{0.4, 0.4}\\
        & DirAug ($p$) \cite{morocutti2023device}& \multicolumn{3}{c|}{0.4}& \multicolumn{2}{c}{0.6}\\
        \midrule
        \multirow{6}{*}{\rotatebox{0}{Training}}& Optimizer& \multicolumn{4}{c|}{Adam}& AdamW\\
        & Learning rate scheduler& \multicolumn{2}{c|}{CosineAnnealingWarmRestarts \cite{loshchilov2017sgdr}}& \multicolumn{3}{c}{WarmUpLinearDown \cite{Schmid2023workshop}}\\
        & Peak learning rate& \multicolumn{2}{c|}{0.004}& 0.001& \multicolumn{2}{c}{1e-5 (Stage 2: 0.001)}\\
        & Max epochs& \multicolumn{2}{c|}{150}& 50& 13& 30\\
        & Warmup, Down& \multicolumn{2}{c|}{-}& 5, 45& 3, 10& 4, 26\\
        & $T_0,T_{\mathrm{mult}}$& \multicolumn{2}{c|}{10, 2}& -& -& -\\
        \bottomrule
    \end{tabular}
    \label{tab:appendix}
\end{table*}

\end{document}